\documentclass[amsmath,amssymb,twocolumn,prd,floatfix,showpacs,nofootinbib]{revtex4-1}

\usepackage{xcolor}          
\usepackage[normalem]{ulem}   
\usepackage[mathlines]{lineno}
\usepackage{xspace}

\usepackage{tabularx}
\usepackage{bm}
\usepackage{overpic,subfigure}
\usepackage{multirow}
\usepackage{array}
\usepackage{dcolumn}
\usepackage[symbol]{footmisc}
\usepackage{epstopdf}

\definecolor{RED}{rgb}{1,0,0}
\definecolor{BLUE}{rgb}{0,0,1}

\usepackage{hyperref}
\hypersetup{
	hypertex=true,
	colorlinks=true,
	linkcolor=blue,
	urlcolor=blue,
	citecolor=blue
}

\newcommand{\BESIIIorcid}[1]{\href{https://orcid.org/#1}{\hspace*{0.1em}\raisebox{-0.45ex}{\includegraphics[width=1em]{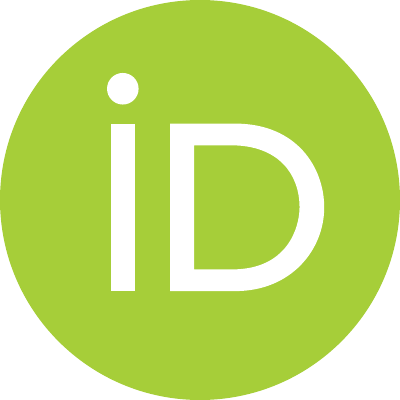}}}}

\begin{document}

\title{\boldmath Observation of $\chi_{cJ}\to \Lambda K^0_S\bar \Xi^0 + c.c.$ and improved measurements of $\chi_{cJ}\to \Lambda K^-\bar \Xi^+ + c.c.$ }

\author{M.~Ablikim \textit{et al.}}
\thanks{Full author list given at the end of the article.}

\collaboration{BESIII Collaboration}
\noaffiliation

\begin{abstract}
  By analyzing $(2712.4\pm14.3)\times10^6$ $\psi(3686)$ events collected with the BESIII detector operating at the BEPCII collider, we report 
  the first observation of $\chi_{cJ}\to \Lambda K^0_S\bar \Xi^0 + c.c.$ ($J=0,1,2$) with significances greater than $10\sigma$, including systematic uncertainties, and improved measurements of $\chi_{cJ}\to \Lambda K^-\bar \Xi^+ + c.c.$
  The branching fractions of these decays are determined to be
$\mathcal{B}(\chi_{c0}\to \Lambda K^0_S\bar \Xi^0+c.c.)=(1.13\pm0.13\pm0.13)\times10^{-4}$,   
$\mathcal{B}(\chi_{c1}\to \Lambda K^0_S\bar \Xi^0+c.c.)=(0.74\pm0.08\pm0.07)\times10^{-4}$,
$\mathcal{B}(\chi_{c2}\to \Lambda K^0_S\bar \Xi^0+c.c.)=(1.16\pm0.11\pm0.14)\times10^{-4}$,
$\mathcal{B}(\chi_{c0}\to \Lambda K^-\bar \Xi^++c.c.)=(1.90\pm0.08\pm0.16)\times10^{-4}$,
$\mathcal{B}(\chi_{c1}\to \Lambda K^-\bar \Xi^++c.c.)=(1.40\pm0.05\pm0.11)\times10^{-4}$, and
$\mathcal{B}(\chi_{c2}\to \Lambda K^-\bar \Xi^++c.c.)=(1.95\pm0.07\pm0.15)\times10^{-4}$.
The first uncertainties are statistical, and the second are systematic.

\end{abstract}

\maketitle

\section{Introduction}

Charmonium lies in the transition region between the perturbative and non-perturbative regimes of quantum chromodynamics (QCD)~\cite{ref::heavyquarksystems,ref::quarkonia}. Its production and decay properties provide important tests of QCD-based models. Notably, the $P$-wave $\chi_{cJ}$ ($J=0,1,2$) states are well suited for investigating the role of color-octet components in charmonium decay mechanisms~\cite{ref::oct1,ref::oct2}. The direct production of $\chi_{cJ}$ states in $e^+e^-$ collisions is highly suppressed~\cite{ref::eetochic1}. They are primarily studied through radiative transitions from higher vector charmonium states, particularly $\psi(3686)\to \gamma\chi_{cJ}$. Consequently, their decay properties remain relatively unexplored. The BESIII and CLEO collaborations have reported measurements of $\chi_{cJ}$ decays into $B_8 \bar{B}_8 M_8$, such as $p\bar{p}\pi^0$, $p\bar\Sigma^+K^0_S$, and $\Lambda\bar\Lambda\eta$ ~\cite{ref::chicj_ppbarX,ref::chicj_sigmapks,ref::chicj_LLbareta}, where $B_8$ and $M_8$ denote the baryon octet and pseudoscalar meson octet, respectively.
These decays allow searches for near-threshold enhancements in baryonic systems, analogous to the enhancement observed in $J/\psi \to \gamma p\bar{p}$~\cite{ref::ppbar1,ref::ppbar2}. In one such study, an enhancement near the $\bar{p}\Lambda$ threshold was observed in $\chi_{cJ}\to \bar{p}K^+\Lambda$~\cite{ref::chicj_pklambda} (charge conjugate modes are implied throughout unless explicitly stated otherwise). This observation suggests that analogous near-threshold structures may also appear in other strange baryon-antibaryon systems.
However, experimental information on the $\Lambda\bar{\Xi}$ system in $\chi_{cJ}$ decays remains limited.

Previously, BESIII measured the branching fractions (BFs) of the decays $\chi_{cJ}\to \Lambda K^-\bar \Xi^+$ using $(107.7\pm0.6)\times10^6$ $\psi(3686)$ events collected in 2009, obtaining
$\mathcal{B}(\chi_{c0}\to \Lambda K^-\bar \Xi^+)=(1.96\pm0.31\pm0.16)\times10^{-4}$,
$\mathcal{B}(\chi_{c1}\to \Lambda K^-\bar \Xi^+)=(1.43\pm0.22\pm0.12)\times10^{-4}$, and
$\mathcal{B}(\chi_{c2}\to \Lambda K^-\bar \Xi^+)=(1.93\pm0.30\pm0.15)\times10^{-4}$~\cite{ref::chicj_LXiK}.
Assuming approximate isospin symmetry, the BFs for
$\chi_{cJ}\to \Lambda K^0_S\bar{\Xi}^0$
are expected to be comparable to those of
$\chi_{cJ}\to \Lambda K^-\bar{\Xi}^+$,
suggesting that this decay mode should be accessible using the current $\psi(3686)$ data sample.
In this paper, we present the first observation of $\chi_{cJ} \to \Lambda K^0_S\bar \Xi^0$ and improved measurements of $\chi_{cJ} \to \Lambda K^-\bar \Xi^+$ by analyzing $(2712.4\pm14.3)\times10^6$ $\psi(3686)$ events~\cite{ref::psip-num-inc} collected with the BESIII detector.

\section{BESIII DETECTOR AND MONTE CARLO SIMULATION}
\label{sec:BES}

The BESIII detector~\cite{ref::detector} records symmetric $e^+e^-$ collisions provided by the BEPCII storage ring~\cite{ref::collider}
at center-of-mass energies ranging from 1.84 to 4.95~GeV,
with a peak luminosity of $1.1 \times 10^{33}\;\text{cm}^{-2}\text{s}^{-1}$
achieved at $\sqrt{s} = 3.773\;\text{GeV}$.

The cylindrical core of the BESIII detector covers 93\% of the full solid angle and consists of a helium-based
 multilayer drift chamber~(MDC), a plastic scintillator time-of-flight
system~(TOF), and a CsI(Tl) electromagnetic calorimeter~(EMC),
which are all enclosed in a superconducting solenoidal magnet
providing a 1.0~T magnetic field.
The solenoid is supported by an
octagonal flux-return yoke with resistive plate counter muon
identification modules interleaved with steel.
The charged-particle momentum resolution at $1~{\rm GeV}/c$ is
$0.5\%$. 
The EMC measures photon energies with a
resolution of $2.5\%$ ($5\%$) at $1$~GeV in the barrel (end cap)
region. The time resolution in the TOF barrel region is 68~ps, while
that in the end cap region was 110~ps.
The end-cap TOF system was upgraded in 2015 using multi-gap resistive plate chamber technology, providing a time resolution of 60~ps~\cite{Tof1,Tof2,Tof3}. Around 83\% of the data used in this analysis benefits from this upgrade.

Simulated data samples, produced with a {\sc geant4}-based~\cite{Geant4} Monte Carlo (MC) package that incorporates the geometric description of the BESIII detector and its response, are used to determine detection efficiencies and estimate background contributions. The simulation models the beam
energy spread and initial state radiation in the $e^+e^-$
annihilations with the generator {\sc
kkmc}~\cite{Jadach01}.

The inclusive MC sample includes the production of the
$\psi(3686)$ resonance, the production of the $J/\psi$ meson via initial-state radiation, and
the continuum processes incorporated in {\sc
kkmc}~\cite{Jadach01}.
All particle decays are modeled with {\sc
evtgen}~\cite{Lange01,Lange02} using BFs
either taken from the
Particle Data Group~(PDG)~\cite{ref::pdg2026}, when available,
or otherwise estimated with {\sc lundcharm}~\cite{ref::Lundcharm0,ref::Lundcharm1}.
Final state radiation
from charged final state particles is incorporated using the {\sc
photos} package~\cite{PHOTOS}.
An inclusive MC sample containing $2.7\times10^{9}$ generic $\psi(3686)$ decays is used to study background. 
For each signal process of $\psi(3686)\to\gamma\chi_{cJ}$ with the final state $\Lambda K\bar{\Xi}$, a mixed MC sample is used to evaluate the detection efficiency.
The $\psi(3686)\to\gamma\chi_{cJ}$ transitions are generated such that the radiative photon follows a ($1+\lambda\cos^{2}\theta$) angular distribution, where $\theta$ is the angle between the photon and the beam direction, and $\lambda = 1$, $-1/3$, and $1/3$ for $J = 0$, $1$, and $2$, respectively. The masses and widths of the $\chi_{cJ}$ mesons are taken from PDG~\cite{ref::pdg2026}.
The $\Lambda K\bar{\Xi}$ final state is modeled as arising from a combination of direct $\Lambda K\bar{\Xi}$ production, simulated with the phase space (PHSP) model, and intermediate resonant processes, including $\chi_{cJ}$ decaying to $\Xi(1690) \bar \Xi$, $\Xi(1820) \bar \Xi$, $\Xi(1910) \bar \Xi$, $\Xi(2030) \bar \Xi$, $\Lambda(1920) \bar \Lambda$, $\Lambda(1950) \bar \Lambda$, $\Lambda(2030) \bar \Lambda$, and $K(2650)^{*+} \bar K^-$,
where the intermediate states subsequently decay to $\Lambda K\bar{\Xi}$. The resonant processes, such as $\chi_{cJ}\to \Xi(1690) \bar \Xi$ and $\Xi(1690)\to\Lambda K^-$, are also generated assuming a uniform population of the available phase space for $n$-body decays, with no spin-dependent effects included. Each mixed MC sample is constructed by combining a subset of these subprocesses and contains $1.0\times10^{6}$ generated events.
The $\Xi(1910)$, $\Xi(2030)$, $\Lambda(1920)$, $\Lambda(1950)$, $\Lambda(2030)$, and $K(2650)^+$ states are included in the MC simulation as possible intermediate contributions for efficiency determination. No attempt is made in this work to confirm or study these states.
The fractions of all intermediate processes are obtained from one-dimensional fits to the invariant-mass spectra of the three two-body combinations, $\Lambda K$, $\Lambda\bar{\Xi}$, and $\bar{\Xi}K$, in data.

\section{EVENT SELECTION}
\label{sec:selection}

We reconstruct the decays $\psi(3686) \to \gamma \chi_{cJ}$ with $\chi_{cJ} \to \Lambda K^0_S\bar \Xi^0$ and $\chi_{cJ} \to \Lambda K^-\bar \Xi^+$, where $\Lambda \to p\pi^-$, $K^0_S \to \pi^+\pi^-$, and the $\bar \Xi^0$ ($\bar \Xi^+$) baryon is reconstructed via $\bar \Xi^0 \to \pi^0 \bar \Lambda$ ($\pi^{0}\to\gamma\gamma$) ($\bar \Xi^+ \to \pi^+ \bar \Lambda$).
Signal events are required to contain at least six charged tracks and at least three (one) photon candidates for $\chi_{cJ} \to \Lambda K^0_S\bar{\Xi}^0$ ($\chi_{cJ} \to \Lambda K^-\bar{\Xi}^+$).
 
 The $K^-$ candidates are required to satisfy $|\cos\theta|<0.93$, $|V_{z}|<10$~cm, and $V_{xy}<1$~cm, where $|V_{z}|$ is the distance of closest approach to the IP along the $z$-axis, and $V_{xy}$ is the distance in the transverse plane. The $\pi^-$ candidates used to reconstruct $K_S^0$, $\Lambda$, and $\Xi$ are required to satisfy $|V_z|<20$~cm and $|\cos\theta|<0.93$. The same requirements are applied to the proton candidates used to reconstruct $\Lambda$.
 
 Particle identification~(PID) for charged tracks combines measurements of the d$E$/d$x$ in the MDC and the flight time in the TOF to form likelihoods $\mathcal{L}(h)$ for each hadron hypothesis $h~(h=p, K, \pi)$.
 The charged tracks with $\mathcal{L}(p)>\mathcal{L}(\pi)$ and $\mathcal{L}(p)>\mathcal{L}(K)$ are assigned as protons, the remaining tracks with $\mathcal{L}(K)>\mathcal{L}(\pi)$ are identified as kaons, and the rest are identified as pions.
 
 The $K^0_S$ candidate is reconstructed from two oppositely charged pions. These tracks are constrained to originate from a common vertex by applying a vertex fit, and the decay length, defined as the distance between the IP and the $K_S^0$ decay vertex, is required to be greater than twice its resolution.
 The $K^{0}_{S}$ signal region in the invariant mass of the $\pi^+ \pi^-$ system $M_{\pi^+ \pi^-}$ is set to be $|M_{\pi^+ \pi^-}-m_{K^{0}_{S}}|<0.012~\mathrm{GeV/}$${c}^2$, where $m_{K^{0}_{S}}$ is the $K^0_S$ nominal mass~\cite{ref::pdg2026}.~The $K^{0}_{S}$ sideband region is set to be $0.020~\mathrm{GeV/}$${c}^2<|M_{\pi^+ \pi^-}-m_{K^{0}_{S}}|<0.044~\mathrm{GeV/}$${c}^2$.

Photon candidates are identified from electromagnetic showers reconstructed in the EMC. The deposited energy of each shower must exceed 25~MeV in the barrel region ($|\cos \theta|< 0.80$, where $\theta$ denotes the
polar angle with respect to the $z$-axis, the symmetry axis of the MDC) and more than 50~MeV in the end cap region ($0.86 <|\cos \theta|< 0.92$). Showers associated with the charged tracks are rejected by requiring the opening angle between the shower and the nearest charged track at the EMC must be greater than 10 degrees as measured from the interaction point (IP). To suppress electronic noise and unrelated showers, the difference between the EMC time and the event start time is required to be within [0,700]\,ns.

The $\pi^{0}$ candidates are reconstructed from photon pairs via the decay $\pi^{0}\to\gamma\gamma$. A photon pair is accepted as a $\pi^{0}$ candidate if its invariant mass lies within $(0.115, 0.150)$~GeV$/c^{2}$, corresponding to approximately $(-4\sigma, 3\sigma)$ around the nominal $\pi^{0}$ mass, to account for the bias in the EMC response to low-energy photons. A one-constraint (1C) kinematic fit with the $\pi^{0}$ mass constraint is then applied to the four-momentum of the two-photon system to improve the momentum resolution. If more than one $\pi^{0}$ candidate is present in an event, the best candidate is selected according to the subsequent $\Xi^{0}$ selection criteria.

The $\Lambda$ candidates are reconstructed from the $\Lambda \to p\pi^-$ decay. The $p\pi^-$ tracks are constrained to originate from a common vertex by applying a vertex fit, and the decay length is required to be greater than twice its resolution.
The $\Lambda$ signal region is set to be $|M_{p\pi^-}-m_{\Lambda}|<0.004~\mathrm{GeV}/c^2$, and the $\Lambda$ sideband region is set to be $0.008<|M_{p \pi^-}-m_{\Lambda}|<0.016~\mathrm{GeV}/c^2$, where $m_{\Lambda}$ is the $\Lambda$ nominal mass~\cite{ref::pdg2026}.

The $\bar \Xi^+$ candidates are reconstructed via the $\bar \Xi^+\to \bar \Lambda \pi^+$ decay, where the $\bar \Lambda$ candidates are required to satisfy the selection criteria described above. No PID requirement is imposed on this track, which is assumed to be a pion. 
A common vertex constraint is further imposed on the $\bar{\Lambda}$ and $\pi^+$ candidates.
If multiple $\bar \Xi^+$ candidates are found, the candidate with the minimum 
$\sqrt{(\frac{M_{\bar \Lambda \pi^+}-m_{\bar{\Xi}^+}}{\sigma_{\bar{\Xi}^+}})^2
	+(\frac{M_{\bar{p}\pi^+}-m_{\bar{\Lambda}}}{\sigma_{\bar{\Lambda}}})^2}$ 
value is retained, where $m_{\bar{\Xi}^+}$ and $m_{\bar{\Lambda}}$ are the $\bar \Xi^+$ and $\bar{\Lambda}$ masses from PDG~\cite{ref::pdg2026}, $\sigma_{\bar{\Xi}^+}$ and $\sigma_{\bar{\Lambda}}$ are the mass resolutions of $\bar\Xi^+$ and $\bar{\Lambda}$, which are 2.3~MeV/$c^2$ and 1.5~MeV/$c^2$ respectively, as determined from MC simulations.
The $\bar \Xi^+$ signal region is set to be $|M_{\bar \Lambda \pi^+}-m_{\bar{\Xi}^+}|<0.006~\mathrm{GeV}/c^2$, and the $\bar \Xi^+$ sideband region is set to be $0.012<|M_{\bar \Lambda \pi^+}-m_{\bar{\Xi}^+}|<0.018~\mathrm{GeV}/c^2$.

The $\bar \Xi^0$ candidates are reconstructed via the decay $\bar \Xi^0\to \bar \Lambda \pi^0$, where $\bar \Lambda$ candidates are selected as described above, and $\pi^0$ candidates are reconstructed according to the corresponding procedure.
If multiple $\bar \Xi^0$ candidates are found, the one that best satisfies the mass constraints on the intermediate states is retained, by minimizing
$\sqrt{ \left(\frac{M_{\bar \Lambda \pi^0}-m_{\bar{\Xi}^0}}{\sigma_{\bar{\Xi}^0}}\right)^2
	+\left(\frac{M_{\bar{p}\pi^+}-m_{\bar{\Lambda}}}{\sigma_{\bar{\Lambda}}}\right)^2
	+\left(\frac{M_{\gamma\gamma}-m_{\pi^0}}{\sigma_{\pi^0}}\right)^2 }
$. Here, $m_{\bar{\Xi}^0}$, $m_{\bar{\Lambda}}$, and $m_{\pi^0}$ are the known masses~\cite{ref::pdg2026}, and $\sigma_{\bar{\Xi}^0}$, $\sigma_{\bar{\Lambda}}$, and $\sigma_{\pi^0}$ denote the corresponding mass resolutions of 5.0~MeV/$c^2$, 1.5~MeV/$c^2$, and 5.0~MeV/$c^2$, respectively.
The $\bar \Xi^0$ signal region is defined as $|M_{\bar \Lambda\pi^0}-m_{\bar{\Xi}^0}|<0.012~\mathrm{GeV}/c^2$, while the sideband region is defined as $0.020<|M_{\bar \Lambda\pi^0}-m_{\bar{\Xi}^0}|<0.044~\mathrm{GeV}/c^2$.

A four-momentum conservation kinematic fit (4C) is applied to each event under the hypothesis of $ e^+e^-\to p\bar{p}\pi^+\pi^-\pi^+\pi^-\pi^0\gamma$ or $ e^+e^-\to p\bar{p}\pi^+\pi^-\pi^+K^-\gamma$. When multiple combinations survive in an event, the one with the smallest 4C fit $\chi^{2}$ ($\chi_{\rm 4C}^{2}$) is retained.

The requirement on $\chi^2_{\rm 4C}$ is optimized using a figure of merit defined as $\mathrm{FOM}\equiv S/\sqrt{S+B}$, where $S$ and $B$ denote the signal and background yields obtained from MC samples, respectively. The signal yield $S$ is defined as mixed signal MC sample, normalized according to the measured BFs, while the background yield $B$ is determined using the inclusive MC sample, normalized to the size of data sample.
Based on the optimization, requirements of $\chi^2_{\rm 4C}<100$ and $\chi^2_{\rm 4C}<70$ are imposed for the $\chi_{cJ}\to\Lambda K_S^0\bar{\Xi}^0$ and $\chi_{cJ}\to\Lambda K^-\bar{\Xi}^+$ selections, respectively.

Potential background contributions from $\psi(3686)$ decays are investigated using the $\psi(3686)$ inclusive MC sample with the generic event type analysis tool TopoAna~\cite{topo}. Only a negligible number of peaking background events are observed in the $M_{\Lambda K \bar{\Xi}}$ distribution, and these are found to be well described by the $\Lambda$, $\Xi$, and $K^0_S$ sideband samples. The continuum background contribution is
evaluated using data taken at $\sqrt{s} = 3.650$~GeV, corresponding to an integrated luminosity of 0.45~fb$^{-1}$~\cite{lum}, and is found to be negligible.

\section{BRANCHING FRACTION}

To determine the signal yields of $\chi_{cJ}\to\Lambda K^0_S\bar{\Xi}^0$, a simultaneous unbinned maximum-likelihood fit is performed to the $M_{\Lambda K^0_S\bar{\Xi}^0}$ distributions of accepted candidates in both the signal and sideband regions of the $\Lambda$, $\bar{\Xi}^0$, and $K^0_S$. The $\chi_{cJ}$ signal shape is modeled as a Breit--Wigner function multiplied by a factor of $E_{\gamma}^3 \cdot D(E_{\gamma})$ and convolved with a Gaussian function, applied consistently in both the signal and sideband regions, with all parameters free except for the width, which is fixed to the PDG value~\cite{ref::pdg2026}. Here, $E_{\gamma} = \frac{m_{\psi(3686)}^2 - m_{\Lambda K_S^0 \bar{\Xi}^0}^2}{2m_{\psi(3686)}}$ is the energy of the transition photon in the $\psi(3686)$ rest frame, and the damping factor $D(E_{\gamma})$ is introduced to suppress the divergent high-energy tail following the prescription of the KEDR collaboration~\cite{ref::damping}: $D^{\mathrm{KEDR}}=\frac{E_0^2}{E_{\gamma}E_0+(E_{\gamma}-E_0)^2}$, where $E_0=\frac{m_{\psi(3686)}^2-m_{\chi_{cJ}}^2}{2m_{\psi(3686)}}$ is the peak energy of the transition photon.
The combinatorial background is described by a second-order polynomial function in both the signal and sideband regions. 
The peaking background contributions from non-$\Lambda$, non-$\bar{\Xi}^0$, and non-$K_S^0$ sources in the $\chi_{cJ}$ signal region are estimated by fitting the corresponding $\Lambda$, $\bar{\Xi}^0$, and $K_S^0$ sideband regions with the same model used for the signal region, except that the resolution is fixed to the value determined from fits to mixed MC samples in the corresponding sideband regions. The resulting yields are then normalized according to the volume ratios of the signal and sideband regions, with contributions from the two- and three-dimensional sidebands subtracted under the assumption of uniform background distribution across the sideband regions, and subsequently fixed in the signal-region fit.

The signal yields of $\chi_{cJ}\to\Lambda K^-\bar{\Xi}^+$ are determined using an analogous simultaneous unbinned maximum-likelihood fit to the $M_{\Lambda K^- \bar{\Xi}^+}$ distributions of candidates in the signal and sideband regions of the $\Lambda$ and $\bar{\Xi}^+$.
The peaking background contributions in the $\chi_{cJ}$ signal region are estimated from events in the $\Lambda$ and $\bar{\Xi}^+$ sideband regions, whose yields are normalized to the signal-to-sideband area ratios, with corrections applied for the  two-dimensional sidebands.
		
\begin{figure}[htbp]
	\centering
	\includegraphics[width=8cm]{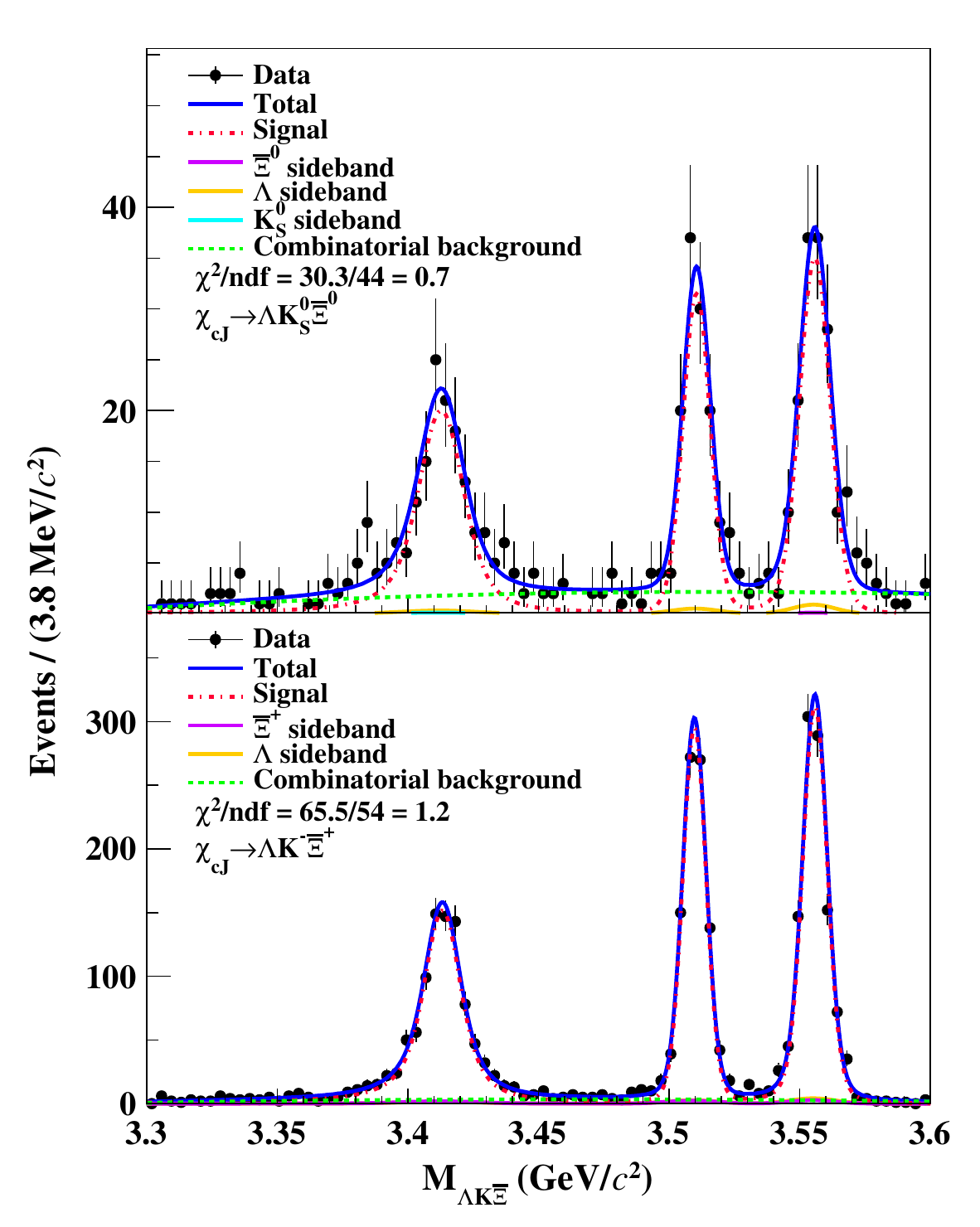}
	\caption{The simultaneous fit to the $M_{\Lambda K \bar \Xi}$ distribution in the $\bar \Xi$, $\Lambda$, and $K^0_S$ signal regions for (top) $\chi_{cJ}\to \Lambda K^0_S \bar \Xi^0$ and (bottom) $\chi_{cJ}\to \Lambda K^- \bar \Xi^+$. The black dots with error bars are data. The blue solid curves are the fit results. The red dashed curves represent the fitted signal components. The purple, orange, and cyan solid curves denote the peaking background contributions constrained by the $\bar \Xi$, $\Lambda$, and $K^0_S$ sideband events in data, respectively. The green dashed curves represent the combinatorial background contribution.}
	\label{tab:LambdaKXi}
\end{figure}

Figure~\ref{tab:LambdaKXi} shows the fit results. The signal yields of $\chi_{c0,1,2}\to\Lambda K^0_S\bar{\Xi}^0$ are determined to be $141\pm15$, $111\pm12$, and $145\pm14$, while those of $\chi_{0,1,2}\to\Lambda K^-\bar{\Xi}^+$ are $912\pm35$, $900\pm32$, and $1052\pm35$, respectively.
The statistical significances of the $\chi_{cJ}\to\Lambda K^0_S\bar{\Xi}^0$ signal candidates all exceed $10\sigma$, as determined from the change in likelihood between the nominal fit and the fit with the signal yields fixed to zero. The significance is evaluated using a Bayesian approach with systematic uncertainties incorporated~\cite{ref::significance1,ref::significance2}.
		
\begin{figure*}[htbp]
	\centering
	\includegraphics[width=13cm]{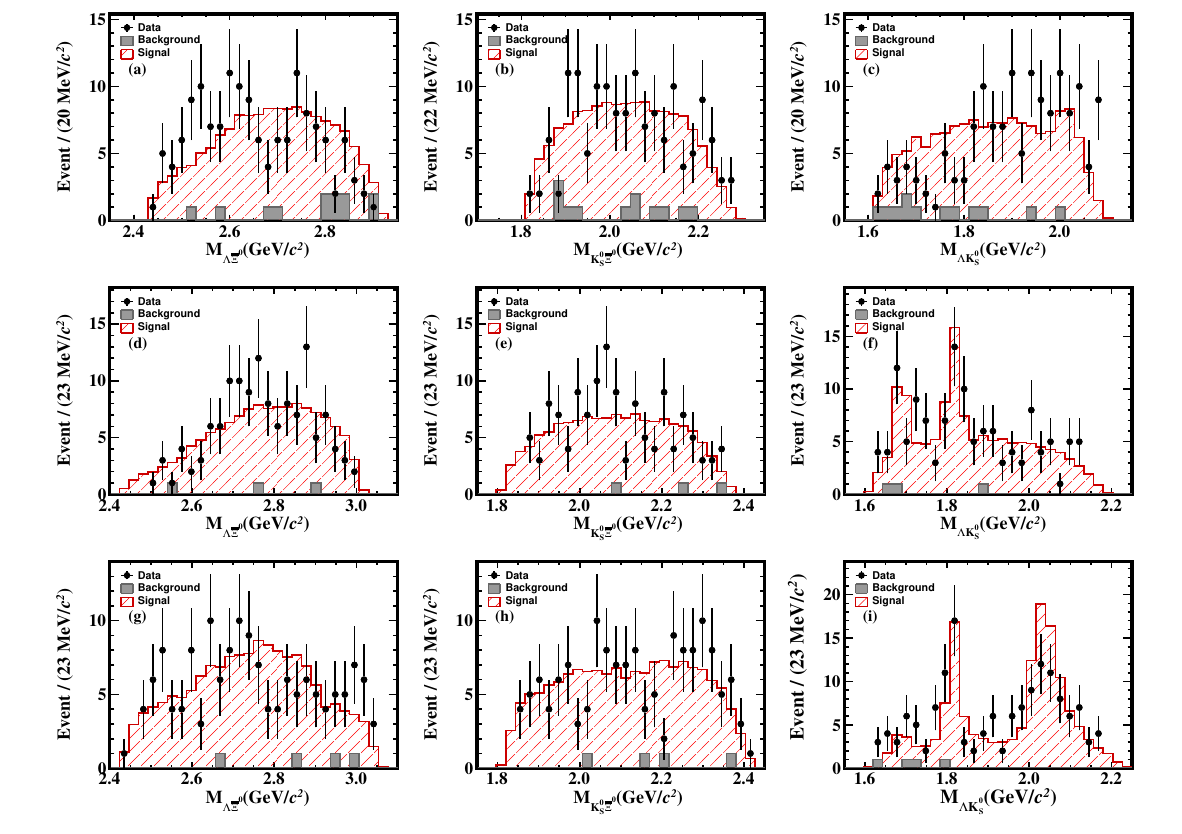}	
	\caption{Comparisons of $M_{\Lambda \bar \Xi^0}$, $M_{K^0_S \bar \Xi^0}$, and $M_{\Lambda K^0_S}$ distributions of the accepted candidates for $\chi_{cJ} \to \Lambda K^0_S \bar{\Xi}^0$. The black dots with error bars represent data, the histograms filled with red lines denote the mixed signal MC sample, and the grey histograms show the background contributions from the inclusive MC sample. Panels~(a)--(c), (d)--(f), and (g)--(i) show the corresponding distributions for events within the $\chi_{c0}$, $\chi_{c1}$, and $\chi_{c2}$ signal regions, respectively, defined by $M_{\Lambda K^0_S \bar{\Xi}^0} \in [3.400, 3.430]$, $[3.501, 3.521]$, and $[3.546, 3.566]$ GeV/$c^2$.}
	\label{fig:compare_LXiKS}
\end{figure*}

Figure~\ref{fig:compare_LXiKS} shows comparisons of the two-body invariant-mass distributions for $\chi_{cJ} \to \Lambda K^0_S\bar{\Xi}^0$ between data and mixed signal MC samples. 
Localized discrepancies between data and MC are observed in the $M_{\Lambda K^0_S}$ distribution around 1.70~GeV/$c^2$ in the $\chi_{c0}$ signal region, which may arise from the limited description of the underlying decay dynamics in the mixed MC samples. Several smaller deviations observed elsewhere are consistent with statistical fluctuations. Overall, satisfactory agreement between data and MC is achieved.

\begin{figure*}[htbp]
	\centering
	\includegraphics[width=13cm]{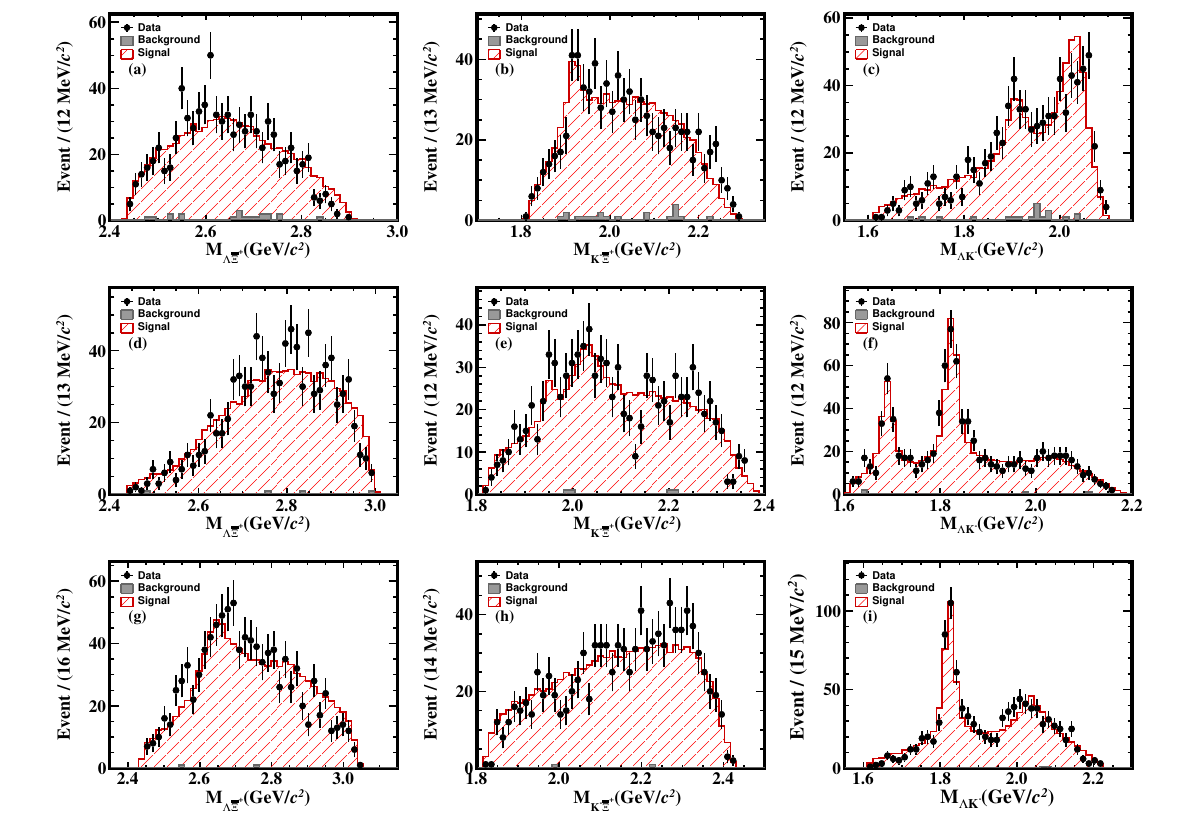}	
	\caption{Comparisons of $M_{\Lambda \bar \Xi^+}$, $M_{K^- \bar \Xi^+}$, and $M_{\Lambda K^-}$ distributions of the accepted candidates for $\chi_{cJ} \to \Lambda K^- \bar{\Xi}^+$. The black dots with error bars represent data, the histograms filled with red lines denote the mixed signal MC sample, and the grey histograms show the background contributions from the inclusive MC sample. Panels~(a)--(c), (d)--(f), and (g)--(i) show the corresponding distributions for events within the $\chi_{c0}$, $\chi_{c1}$, and $\chi_{c2}$ signal regions, respectively, defined by $M_{\Lambda K^- \bar{\Xi}^+} \in [3.400, 3.430]$, $[3.501, 3.521]$, and $[3.546, 3.566]$ GeV/$c^2$.}
	\label{fig:compare_LXiK}
\end{figure*}

Figure~\ref{fig:compare_LXiK} shows the corresponding comparisons for $\chi_{cJ} \to \Lambda K^-\bar{\Xi}^+$. Localized discrepancies are observed in the $M_{K^-\bar{\Xi}^+}$ distribution around 2.13~GeV/$c^2$ in the $\chi_{c1}$ signal region, which may also originate from the limited modeling of the decay dynamics in the mixed MC samples.

In the $\Lambda K$ invariant-mass distributions, clear structures consistent with the $\Xi(1690)$ and $\Xi(1820)$ are observed in the charged mode, along with a possible structure around 1.91~GeV/$c^2$. Indications of the $\Xi(1690)$ and $\Xi(1820)$ are also seen in the neutral mode. No significant enhancement is observed near the $\Lambda\bar{\Xi}$ mass threshold. The detection efficiencies for $\chi_{cJ} \to \Lambda K^0_S\bar{\Xi}^0$ and $\chi_{cJ} \to \Lambda K^-\bar{\Xi}^+$ are listed in Table~\ref{tab:BFs}.

The product of BFs are calculated as
\begin{linenomath*}
	\begin{equation}
		{\mathcal B}(\psi(3686)\to \gamma\chi_{cJ}) \cdot \mathcal{B}(\chi_{cJ}\to \Lambda K \bar \Xi)
		=\frac{N^{\rm obs}_{\chi_{cJ}}}{N_{\psi(3686)}\cdot\epsilon \cdot \mathcal B_{\rm sub}},
	\end{equation}
\end{linenomath*}
where $N^{\rm obs}_{\chi_{cJ}}$ is the signal decay yield extracted from the fit,
$N_{\psi(3686)}$ denotes the total number of $\psi(3686)$ events in data, and $\epsilon$ is the corresponding detection efficiency. The variable $\mathcal B_{\rm sub}$ represents the product of BFs of the intermediate decay modes: $\mathcal B(K^0_S \to \pi^+ \pi^-) \cdot \mathcal B^2(\Lambda  \to p \pi^-) \cdot \mathcal B(\bar\Xi^0  \to\bar \Lambda \pi^0)
\cdot \mathcal B(\pi^0 \to \gamma\gamma)$ for $\chi_{cJ} \to \Lambda K^0_S \bar \Xi^0$, and
$\mathcal B^2(\Lambda  \to p \pi^-) \cdot \mathcal B(\bar\Xi^+  \to\bar \Lambda \pi^-)$
for $\chi_{cJ} \to \Lambda K^-\bar \Xi^+$.
The BFs of $\mathcal{B}(\chi_{cJ}\to \Lambda K \bar \Xi)$ are then obtained by dividing the measured products by the $\psi(3686)\to\gamma\chi_{cJ}$ BF taken from the PDG~\cite{ref::pdg2026}. All results for $\chi_{cJ}\to \Lambda K^0_S \bar \Xi^0$ and $\chi_{cJ}\to \Lambda K^- \bar \Xi^+$ are summarized in Tables \ref{tab:BFs}.

\section{SYSTEMATIC UNCERTAINTY}
		
The systematic uncertainties in the BF measurements arise from various sources, as summarized in Table~\ref{tab:systematic}, and are discussed below.
		
The total number of $\psi(3686)$ events is determined to be $N_{\psi(3686)}=(2712.4\pm14.3)\times10^6$ using inclusive hadronic events as described in Ref.~\cite{ref::psip-num-inc}, with a corresponding uncertainty of 0.5\%.
	
The systematic uncertainties associated with $K^{-}$ tracking and PID for $K^{-}$ and $p$ are evaluated using control samples of $J/\psi\to K^{*}(892)K_S^0$, $J/\psi\to K^{+}K^{-}\pi^0$, and $J/\psi \to p\bar p \pi^+\pi^-$, respectively, and are assigned as 1.0\% per track for both tracking and PID~\cite{ref::tracking,ref::p_PID}.

The systematic uncertainty of the $K^0_S$ reconstruction, including tracking, mass window range, vertex fit, and secondary vertex fit, is estimated using control samples of $J/\psi \rightarrow K^{*\pm} ¯\bar{K}^\mp$ and $J/\psi \rightarrow \phi K^{*\pm} ¯\bar{K}^\mp$, and is assigned as 1.5\%~\cite{ref::kso}.

The systematic uncertainty of $\Lambda$ reconstruction, covering the tracking of the $p\pi^-$ pair, decay length requirement, mass window requirement, vertex fit and secondary vertex fit, is studied using control samples of $J/\psi\to pK^-\bar \Lambda$ and $J/\psi\to\Lambda\bar \Lambda$. The efficiencies obtained in two-dimensional (2D) intervals are weighted by the 2D distributions of momentum and $\cos\theta$ to match the signal MC events, following the procedure of Ref.~\cite{ref::chicj_LLbareta}. The differences in the re-weighted efficiencies between data and MC simulation are taken as the systematic uncertainties, which are 2.9\%, 1.6\%, 3.5\% for $\chi_{c0,1,2}\to \Lambda K^0_S\bar \Xi^0$, and
2.9\%, 2.5\%, and 2.5\% for $\chi_{c0,1,2}\to \Lambda K^-\bar \Xi^+$, respectively.

The systematic uncertainty of $\bar\Xi^+$ reconstruction is estimated to be $5.1\%$, accounting for the tracking and PID of the $\bar{\Lambda}\pi^+$ pair, $\Lambda$ reconstruction, decay length requirement, mass window range, vertex fit, and secondary vertex fit, using the control sample of $\psi(3686)\to\Xi^-\bar{\Xi}^+$ following Ref.~\cite{ref::Xi-}. The systematic uncertainty of $\bar\Xi^0$ reconstruction is estimated to be $4.5\%$, including the tracking and PID of the $\bar{\Lambda}\pi^0$ pair, $\pi^0$ reconstruction, decay length requirement, and mass window selection, using the control sample of $\psi(3686)\to\Xi^0\bar{\Xi}^0$, according to Ref.~\cite{ref::Xi0}.
		
The systematic uncertainty associated with photon reconstruction is assigned as 1.0\% per photon, based on studies of the control sample of $J/\psi\to\pi^+\pi^-\pi^0$~\cite{ref::gamma-recon}.

The systematic uncertainty from the 4C kinematic fit is evaluated by comparing the signal efficiencies before and after applying a helix parameter correction, described in Ref.~\cite{ref::helixp}.
The resulting changes in signal efficiency are taken as the systematic uncertainties, which are 3.2\%, 1.1\%, and 2.9\% for $\chi_{c0,1,2}\to \Lambda K^0_S\bar \Xi^0$, and 1.9\%, 0.2\%, and 0.3\% for $\chi_{c0,1,2}\to \Lambda K^-\bar \Xi^+$, respectively.

The systematic uncertainties associated with the fits to the $M_{\Lambda K \bar{\Xi}}$ distributions are evaluated by considering several sources: the signal shape, background shape, fitting range, normalization between the signal and sideband regions, possible peaking backgrounds, and photon miscombination.
The uncertainty on the signal shape is assessed by repeating the fits with the width varied within $\pm1\sigma$. The uncertainty on the background shape is evaluated by varying the order of the polynomial used to describe the background. The uncertainty associated with the fitting range is assessed by repeating the fits with ranges varied from [3.28,3.58] to [3.35,3.65]~GeV/$c^2$.
The uncertainty on the normalization between the signal and sideband regions is evaluated by varying the scaling factors according to the ratio of event yields in the two regions, determined from fits to the $M_{\Lambda\pi^0}$, $M_{\Lambda\pi^-}$, $M_{p\pi^-}$, and $M_{\pi^+\pi^-}$ distributions in data.
The systematic uncertainty due to peaking backgrounds from $\chi_{cJ}\to\Sigma^0K\bar{\Xi}$ decays is evaluated by including the MC-simulated component as an additional probability density function (PDF) in the fit, with the yields and relative fractions among the $\chi_{cJ}$ states fixed.
The systematic uncertainty from photon miscombination in $\chi_{cJ}\to\Lambda K_S^0\bar{\Xi}^0$ decays is evaluated by incorporating the miscombined MC shapes into the signal PDF, with their relative fractions fixed according to the miscombination rates determined from a truth-matched MC study.
For each source, the difference between the refitted and nominal signal yields is taken as the corresponding systematic uncertainty. The total fit-related uncertainty, obtained by adding all contributions in quadrature, is 7.7\%, 7.1\%, and 9.4\% for $\chi_{c0,1,2}\to \Lambda K^0_S\bar \Xi^0$, and 2.3\%, 2.0\%, and 2.9\% for $\chi_{c0,1,2}\to \Lambda K^-\bar \Xi^+$, respectively.

The systematic uncertainties associated with fit bias are estimated through input-output checks based on 50 pseudo-data samples constructed from the inclusive MC sample.
If the mean deviation between the fitted and input values exceeds twice its standard deviation, the observed deviation is taken as the systematic uncertainty. These uncertainties are 3.0\%, 3.0\%, and 1.0\% for  $\chi_{c1}\to \Lambda K^0_S\bar \Xi^0$, $\chi_{c2}\to \Lambda K^0_S\bar \Xi^0$, and $\chi_{c2}\to \Lambda K^-\bar \Xi^+$, respectively, and are negligible for the remaining decay modes.

The systematic uncertainties due to the limited statistics of the mixed signal MC samples
are 0.7\%, 0.6\%, and 0.7\% for  $\chi_{c0,1,2}\to  \Lambda K^0_S\bar \Xi^0$,
and 0.5\%, 0.4\%, and 0.4\% for $\chi_{c0,1,2}\to \Lambda K^-\bar \Xi^+$, respectively.

The systematic uncertainties associated with variations in the intermediate resonance fractions of the MC model for $\chi_{cJ}\to \Lambda K\bar \Xi$ decays are evaluated by comparing the nominal signal efficiencies with those obtained from mixed signal MC samples after varying the relative fractions of the sub-resonant decays, namely $\chi_{cJ}\to \Xi(1690)\bar{\Xi}$, $\chi_{cJ}\to \Xi(1820)\bar{\Xi}$, and $\chi_{cJ}\to \Xi(2030)\bar{\Xi}$, by $\pm 1\sigma$. The resulting differences in signal efficiencies, 3.2\%, 0.2\%, and 0.9\%, are assigned as the systematic uncertainties for $\chi_{c0,1,2}\to\Lambda K^0_S\bar \Xi^0$, while the corresponding values of 3.5\%, 0.1\%, and 0.9\% are assigned for $\chi_{c0,1,2}\to\Lambda K^-\bar \Xi^+$, respectively.

		\begin{table*}[htbp]
	\caption{Quantities used in the BFs calculations for $\chi_{cJ} \to \Lambda K^0_S\bar \Xi^0$ and $\chi_{cJ} \to \Lambda K^-\bar \Xi^+$, where the first and second uncertainties are statistical and systematic, respectively.}
	\centering
		\begin{tabular*}{\textwidth}{@{\extracolsep{\fill}}lccc}
			\hline\hline
			& $ \chi_{c0}\to \Lambda K^0_S\bar{\Xi}^0$ & $ \chi_{c1}\to \Lambda K^0_S\bar{\Xi}^0$ & $ \chi_{c2}\to \Lambda K^0_S\bar{\Xi}^0 $ \\
			\hline
			$N_{\rm sig}$ & $141\pm15$ & $111\pm12$ & $145\pm14$ \\
			$\epsilon~(\%)$ & $1.637\pm0.013$ & $2.015\pm0.014$ & $1.724\pm0.013$ \\
			$\mathcal{B}(\psi(3686)\to \gamma \chi_{cJ})\cdot\mathcal{B}(\chi_{cJ}\to \Lambda K^0_S\bar \Xi^0)~(10^{-5})$ & $1.13\pm0.12\pm0.13$ & $0.73\pm0.08\pm0.07$ & $1.11\pm0.11\pm0.14$ \\
			$\mathcal{B}(\chi_{cJ}\to \Lambda K^0_S\bar \Xi^0)~(10^{-4})$ & $1.13\pm0.13\pm0.13$ & $0.74\pm0.08\pm0.07$ & $1.16\pm0.11\pm0.14$ \\ \hline \hline
			& $ \chi_{c0} \to \Lambda K^-\bar{\Xi}^+$ & $ \chi_{c1} \to \Lambda K^-\bar{\Xi}^+$ & $ \chi_{c2} \to \Lambda K^-\bar{\Xi}^+ $ \\
			\hline
			$N_{\rm sig}$ & $912\pm35$ & $900\pm32$ & $1052\pm35$ \\
			$\epsilon~(\%)$ & $4.281\pm0.021$ & $5.870\pm0.024$ & $5.033\pm0.022$ \\
			$\mathcal{B}(\psi(3686)\to \gamma \chi_{cJ})\cdot\mathcal{B}(\chi_{cJ}\to \Lambda K^-\bar \Xi^+)~(10^{-5})$ & $1.91\pm0.07\pm0.16$ & $1.38\pm0.05\pm0.10$ & $1.88\pm0.06\pm0.14$ \\
			$\mathcal{B}(\chi_{cJ}\to \Lambda K^-\bar \Xi^+)~(10^{-4})$ & $1.90\pm0.08\pm0.16$ & $1.40\pm0.05\pm0.11$ & $1.95\pm0.07\pm0.15$ \\ \hline\hline
			$\mathcal{B}_{\Lambda \bar{\Xi}^0K_S^0}/\mathcal{B}_{\Lambda \bar{\Xi}^+K^-}$ & $0.59\pm0.07\pm0.06$ & $0.53\pm0.06\pm0.07$ & $0.59\pm0.06\pm0.06$ \\ \hline
			\hline
		\end{tabular*} 
		\label{tab:BFs}
	\end{table*}
	
	\begin{table*}[htbp]
		\centering
		\caption{Relative systematic uncertainties (in \%) in the BF measurements of $\chi_{cJ} \to \Lambda K^0_S \bar \Xi^0$ and $\chi_{cJ} \to \Lambda K^- \bar \Xi^+$.}
		\begin{tabular*}{0.9\textwidth}{@{\extracolsep{\fill}}lcccccc}
			\hline\hline
			&\multicolumn{3}{c}{$\Lambda K^0_S \bar \Xi^0$}&\multicolumn{3}{c}{$\Lambda K^- \bar \Xi^+$} \\ \hline
			Source & $\chi_{c0}$ & $\chi_{c1}$ & $\chi_{c2}$ & $\chi_{c0}$ & $\chi_{c1}$ & $\chi_{c2}$ \\ \hline
			$N_{\psi(3686)}$ & 0.5 & 0.5 & 0.5 & 0.5 & 0.5 & 0.5\\
			$K^-$ tracking & ... & ... & ... & 1.0 & 1.0 & 1.0\\
			$K^-$ or $p$ PID & 2.0 & 2.0 & 2.0 & 3.0 & 3.0 & 3.0 \\
			$K_{S}^{0}$ reconstruction & 1.5 & 1.5 & 1.5 & ... & ... & ...\\
			$\Lambda$ reconstruction & 2.9 & 1.6 & 3.5 & 2.9 & 2.5 & 2.5\\
			$\bar\Xi^+$ reconstruction & ... & ... & ... & 5.1 & 5.1 & 5.1 \\
			$\bar\Xi^0$ reconstruction & 4.5 & 4.5 & 4.5 & ... & ... & ... \\
			$\gamma$ reconstruction & 1.0 & 1.0 & 1.0 & 1.0 & 1.0 & 1.0 \\
			4C kinematic fit & 3.2 & 1.1 & 2.9 & 1.9 & 0.2  & 0.3\\
			$M_{\Lambda K \bar \Xi}$ fit  & 7.7 & 7.1 & 9.4 & 2.3 & 2.0 & 2.9\\
			Fitting bias & neg. & 3.0 & 3.0 & neg. & neg.  & 1.0\\
			MC statistics & 0.7 & 0.6 & 0.7 & 0.5 & 0.4 & 0.4 \\
			MC generator  & 3.2 & 0.2 & 0.9 & 3.5 & 0.1 & 0.9\\
			$\mathcal B(K_{S}^{0} \to \pi^{+} \pi^{-})$ & 0.1 & 0.1 & 0.1 & ... & ... & ...\\
			$\mathcal B(\pi^0 \to \gamma\gamma )$ & neg. & neg. & neg. & ... & ... & ...\\
			$\mathcal B(\Lambda \to p\pi^-)$ & 1.6 & 1.6 & 1.6  & 1.6 & 1.6 & 1.6 \\
			$\mathcal B(\bar \Xi^0\to \bar \Lambda\pi^0)$ & neg. & neg. & neg. & ... & ... & ...\\
			$\mathcal B(\bar \Xi^+\to \bar \Lambda\pi^+)$ & ... & ... & ... & neg. & neg. & neg. \\
			$\mathcal B(\psi(3686)\to \gamma \chi_{cJ})$ & 2.7 & 2.6 & 2.4 & 2.7 & 2.6 & 2.4 \\
			\hline
			Total~(without $\mathcal B(\psi(3686) \to \gamma\chi_{cJ})$)  & 10.9 & 9.1 & 11.9 & 8.0 & 7.4 & 7.1\\
			Total~(with $\mathcal B(\psi(3686) \to \gamma\chi_{cJ})$)  & 11.2 & 9.5 & 12.1 & 8.4 & 7.9 & 7.5 \\
			\hline\hline
		\end{tabular*}
		\label{tab:systematic}
	\end{table*}
	
The uncertainties from the BFs from the PDG~\cite{ref::pdg2026} are 2.7\%, 2.6\%, and 2.4\% for $\psi(3686)\to \gamma\chi_{c0,1,2}$;
0.07\% for $\mathcal{B}(K^{0}_{S} \to \pi^+ \pi^-)$; 0.8\% for $\mathcal{B}(\Lambda \to p \pi^-)$; 0.01\% for $\mathcal{B}(\bar{\Xi}^0 \to \Lambda\pi^0)$; 0.03\% for $\mathcal{B}(\bar{\Xi}^- \to \Lambda \pi^-)$; and 0.03\% for $\mathcal{B}(\pi^0 \to \gamma \gamma)$.

For each signal decay, the total systematic uncertainty is obtained by adding all individual contributions in quadrature, as listed in Table~\ref{tab:systematic}.

\section{Summary}

Using $(2712.4\pm14.3)\times10^6$ $\psi(3686)$ events collected with the BESIII detector operating at the BEPCII collider,
the hadronic decays $\chi_{cJ}\to \Lambda K^0_S\bar \Xi^0$ are observed for the first time with statistical significances exceeding $10\sigma$,
while the BFs of $\chi_{cJ}\to \Lambda K^-\bar \Xi^+$ are measured with improved precision.
Their branching fractions are determined to be
$\mathcal{B}(\chi_{c0}\to \Lambda K^0_S\bar \Xi^0)=(1.13\pm0.13\pm0.13)\times10^{-4}$,   
$\mathcal{B}(\chi_{c1}\to \Lambda K^0_S\bar \Xi^0)=(0.74\pm0.08\pm0.07)\times10^{-4}$,
$\mathcal{B}(\chi_{c2}\to \Lambda K^0_S\bar \Xi^0)=(1.16\pm0.11\pm0.14)\times10^{-4}$,
$\mathcal{B}(\chi_{c0}\to \Lambda K^-\bar \Xi^+)=(1.90\pm0.08\pm0.16)\times10^{-4}$,
$\mathcal{B}(\chi_{c1}\to \Lambda K^-\bar \Xi^+)=(1.40\pm0.05\pm0.11)\times10^{-4}$, and
$\mathcal{B}(\chi_{c2}\to \Lambda K^-\bar \Xi^+)=(1.95\pm0.07\pm0.15)\times10^{-4}$.
The BF ratios ${\mathcal B}(\chi_{cJ}\to\Lambda K^0_S\bar{\Xi}^0)/{\mathcal B}(\chi_{cJ}\to\Lambda K^-\bar{\Xi}^+)$, listed in Table~\ref{tab:BFs}, are consistent with the theoretical expectation of $0.5$ within uncertainties~\cite{ref::pdg2026}, supporting the validity of isospin symmetry in these decays. In these ratio calculations, the systematic uncertainties associated with the total number of $\psi(3686)$ events, proton PID, $\Lambda$ reconstruction, the radiative photon from $\psi(3686)$, and the external BFs $\mathcal B(\psi(3686)\to \gamma \chi_{cJ})$ and $\mathcal B(\Lambda \to p\pi^-)$ cancel. No threshold enhancements are observed in the $\Lambda \bar \Xi$ mass spectra within the current statistical precision.

\section{Acknowledgement}

The BESIII Collaboration thanks the staff of BEPCII {\hypersetup{urlcolor=black}\href{https://cstr.cn/31109.02.BEPC}{(https://cstr.cn/31109.02.BEPC)}} and the IHEP computing center for their strong support. This work is supported in part by National Key R\&D Program of China under Contracts Nos. 2025YFA1613900, 2023YFA1606000, 2023YFA1606704; National Natural Science Foundation of China (NSFC) under Contracts Nos. 12675104, 11635010, 11935015, 11935016, 11935018, 12025502, 12035009, 12035013, 12061131003, 12192260, 12192261, 12192262, 12192263, 12192264, 12192265, 12221005, 12225509, 12235017, 12342502, 12361141819, 12535005; the Chinese Academy of Sciences (CAS) Large-Scale Scientific Facility Program; the Strategic Priority Research Program of Chinese Academy of Sciences under Contract No. XDA0480600; CAS under Contract No. YSBR-101; 100 Talents Program of CAS; The Institute of Nuclear and Particle Physics (INPAC) and Shanghai Key Laboratory for Particle Physics and Cosmology; Agencia Nacional de Investigación y Desarrollo de Chile (ANID), Chile under Contract No. ANID CCTVal CIA250027; ERC under Contract No. 758462; German Research Foundation DFG under Contract No. FOR5327; Istituto Nazionale di Fisica Nucleare, Italy; Knut and Alice Wallenberg Foundation under Contracts Nos. 2021.0174, 2021.0299, 2023.0315; Ministry of Development of Turkey under Contract No. DPT2006K-120470; National Research Foundation of Korea under Contract No. RS-2026-25486791; National Science and Technology fund of Mongolia; Polish National Science Centre under Contract No. 2024/53/B/ST2/00975; STFC (United Kingdom); Swedish Research Council under Contract No. 2019.04595; U. S. Department of Energy under Contract No. DE-FG02-05ER41374.

\nolinenumbers
\onecolumngrid

\centering
\vspace{0.5cm}
\noindent\rule{0.5\textwidth}{1pt}
\vspace{0.5cm}

M.~Ablikim$^{1}$\BESIIIorcid{0000-0002-3935-619X},
M.~N.~Achasov$^{4,c}$\BESIIIorcid{0000-0002-9400-8622},
P.~Adlarson$^{84}$\BESIIIorcid{0000-0001-6280-3851},
X.~C.~Ai$^{90}$\BESIIIorcid{0000-0003-3856-2415},
C.~S.~Akondi$^{32A,32B}$\BESIIIorcid{0000-0001-6303-5217},
R.~Aliberti$^{40}$\BESIIIorcid{0000-0003-3500-4012},
A.~Amoroso$^{83A,83C}$\BESIIIorcid{0000-0002-3095-8610},
Q.~An$^{79,66,\dagger}$,
Y.~H.~An$^{90}$\BESIIIorcid{0009-0008-3419-0849},
M.~S.~Anderson$^{40}$\BESIIIorcid{0009-0008-1550-2632}
% ...
\\
\vspace{0.2cm}
(BESIII Collaboration)\\
\vspace{0.2cm} {\it
$^{1}$ Institute of High Energy Physics, Beijing 100049, People's Republic of China\\
% ...
$^{4}$ Budker Institute of Nuclear Physics SB RAS (BINP), Novosibirsk 630090, Russia\\
$^{5}$ Carnegie Mellon University, Pittsburgh, Pennsylvania 15213, USA\\
% ...
$^{32}$ INFN Sezione di Ferrara, (A)INFN Sezione di Ferrara, I-44122, Ferrara, Italy; (B)University of Ferrara, I-44122, Ferrara, Italy\\
% ...
$^{40}$ Johannes Gutenberg University of Mainz, Johann-Joachim-Becher-Weg 45, D-55099 Mainz, Germany\\
% ...
$^{66}$ State Key Laboratory of Particle Detection and Electronics, Beijing 100049, Hefei 230026, People's Republic of China\\
% ...
$^{79}$ University of Science and Technology of China, Hefei 230026, People's Republic of China\\
% ...
$^{83}$ University of Turin and INFN, (A)University of Turin, I-10125, Turin, Italy; (B)University of Eastern Piedmont, I-15121, Alessandria, Italy; (C)INFN, I-10125, Turin, Italy\\
$^{84}$ Uppsala University, Box 516, SE-75120 Uppsala, Sweden\\
% ...
$^{90}$ Zhengzhou University, Zhengzhou 450001, People's Republic of China\\
\vspace{0.2cm}
$^{\dagger}$ Deceased\\
% ...
$^{c}$ Also at the Novosibirsk State University, Novosibirsk, 630090, Russia\\
% ...
}

\end{document}